\documentclass[letterpaper,twocolumn,showpacs,
prl %prb
,superscriptaddress
,email
]{revtex4}

\usepackage[pdftex]{graphicx}
\usepackage{amssymb}
\usepackage{amsmath,amsfonts,latexsym}
\usepackage{array,tabularx,color}
\usepackage{dcolumn}               % Align table columns on decimal point
\usepackage[normalem]{ulem}
\usepackage{comment}
\usepackage{bm}

\newcommand{\vect}[1]{{\mathbf #1}}
\begin{document}

%\linenumbers

\title{Onset and dynamics of vortex-antivortex pairs in polariton OPO
  superfluids}

\author{G. Tosi}
\affiliation{F\'isica de Materiales, Universidad
  Aut\'onoma de Madrid, Madrid 28049, Spain}

\author{F. M. Marchetti}
\email{francesca.marchetti@uam.es}
\affiliation{F\'isica Te\'orica de la Materia Condensada, Universidad Aut\'onoma de Madrid, Madrid 28049, Spain}

\author{D. Sanvitto} 
\affiliation{NNL, Istituto Nanoscienze -- CNR, Via Arnesano, 73100 Lecce, Italy}

\author{C. Ant\'on}
\affiliation{F\'isica de Materiales, Universidad Aut\'onoma de Madrid, Madrid 28049, Spain}

\author{M. H. Szyma\'nska}
%\email{M.H.Szymanska@warwick.ac.uk}
\affiliation{Department of Physics, University of Warwick, Coventry,
  CV4 7AL, UK}
\altaffiliation{also at London Centre for Nanotechnology, UK}

\author{A. Berceanu}
\affiliation{F\'isica Te\'orica de la Materia
Condensada, Universidad Aut\'onoma de Madrid, Madrid 28049, Spain}

\author{C. Tejedor}
\affiliation{F\'isica Te\'orica de la Materia
  Condensada, Universidad Aut\'onoma de Madrid, Madrid 28049, Spain}

\author{L. Marrucci}
\affiliation{Dipartimento di Scienze Fisiche, Universit\`a di
Napoli Federico II and CNR-SPIN, Napoli, Italy}

\author{A. Lema\^{i}tre}
\affiliation{LPN/CNRS, Route de Nozay, 91460, Marcoussis, France}

\author{J. Bloch}
\affiliation{LPN/CNRS, Route de Nozay, 91460, Marcoussis, France}

\author{L. Vi\~na}
\affiliation{F\'isica de Materiales, Universidad
  Aut\'onoma de Madrid, Madrid 28049, Spain}

%\date{\today}
\date{May 27, 2011}       % to fix in the last version!

\begin{abstract}
  We study, both theoretically and experimentally, the occurrence of
  topological defects in polariton superfluids in the optical
  parametric oscillator (OPO) regime. We explain in terms of local
  supercurrents the deterministic behaviour of both onset and dynamics
  of vortex-antivortex pairs generated by perturbing the system with a
  pulsed probe. Using a generalised Gross-Pitaevskii equation,
  including photonic disorder, pumping and decay, we elucidate the
  reason why topological defects form in couples and can be detected
  by direct visualizations in multi-shot OPO experiments.
\end{abstract}


%42.65.Yj   Optical parametric oscillators and amplifiers (see also
%           42.65.Lm Parametric down conversion and production of
%           entangled photons)

%47.32.-y   Vortex dynamics; rotating fluids (for vortices in superfluid
%           helium, see 67.25.dk and 67.30.he)

%47.32.C-   Vortex dynamics
 
%47.32.cb   Vortex interactions
 
%47.32.cd   Vortex stability and breakdown

%47.37.+q   Hydrodynamic aspects of superfluidity; quantum fluids 

%03.75.Kk   Dynamic properties of condensates; collective and
%           hydrodynamic excitations, superfluid flow

%03.75.Nt   Other Bose-Einstein condensation phenomena

%42.50.Fx   Cooperative phenomena in quantum optical systems

%67.85.De   Dynamic properties of condensates; 
%           excitations, and superfluid flow 

%71.35.Lk   Collective effects (Bose effects, phase space
%           filling, and excitonic phase transitions)

%71.36.+c   Polaritons (including photon-phonon and
%           photon-magnon interactions)

%71.45.-d   Collective effects

%\keywords{Suggested keywords} %Use showkeys class option if keyword
                               %display desired
\maketitle

%%%%%%%%%%%%%%%%%%%%%%%%%%%%%%%%%%%%%%%
%Introduction
%%%%%%%%%%%%%%%%%%%%%%%%%%%%%%%%%%%%%%%
%
Quantum vortices are topological defects occurring in macroscopically
coherent systems. Their existence was first predicted in
superfluids~\cite{onsager_49,feynman_55}, and later in coherent
waves~\cite{nye74}. Nowadays, quantum vortices have been the subject
of extensive research across several areas of physics and have been
observed in type-II superconductors, ${}^4$He, ultracold atomic gases,
nonlinear optics media (for a review see,
e.g.,~\cite{staliunas,desyatnikova05}) and very recently microcavity
polaritons~\cite{lagoudakis08,lagoudakis09a,lagoudakis10,sanvitto10,krizhanovskii10,roumpos10,nardin11,sanvitto11}.
The phase of a quantised vortex winds around its core from $0$ to
$2\pi m$ (with $m$ integer), implying the vortex carries a quantised
angular momentum, $\hbar m$.
In contrast with the classical counterpart, quantum vortices with the
same $m$ are all identical, with a size (or healing length) determined
by the system nonlinear properties.

Recently, the study of quantized vortices imprinted in polariton
condensates using pulsed laser fields has attracted noticeable
interest both experimentally~\cite{sanvitto10} and
theoretically~\cite{wouters10,marchetti10,szymanska10,gorbach10},
providing a diagnostics for superfluid properties of such a
non-equilibrium system.
In particular, resonantly pumped polaritons in the OPO
regime~\cite{stevenson00,baumberg00:prb} have been recently shown to
exhibit a new form of non-equilibrium
superfluidity~\cite{amo09,sanvitto10}. Here, polaritons continuously
injected into the \emph{pump} state, undergo coherent stimulated
scattering into the \emph{signal} and \emph{idler} states.  An
additional pulsed probe can initiate a traveling decaying gain, which
evolves freely from the probe constraints. By using a pulsed
Laguerre-Gauss (LG) beam, vorticity has been shown to persist not only
in absence of the rotating drive, but also longer than the gain
induced by the probe, and therefore to be transferred to the OPO
signal, demonstrating metastability of quantum vortices and
persistence of currents~\cite{sanvitto10,marchetti10}.

\begin{figure}
\begin{center}
\includegraphics[width=0.65\linewidth,angle=0]{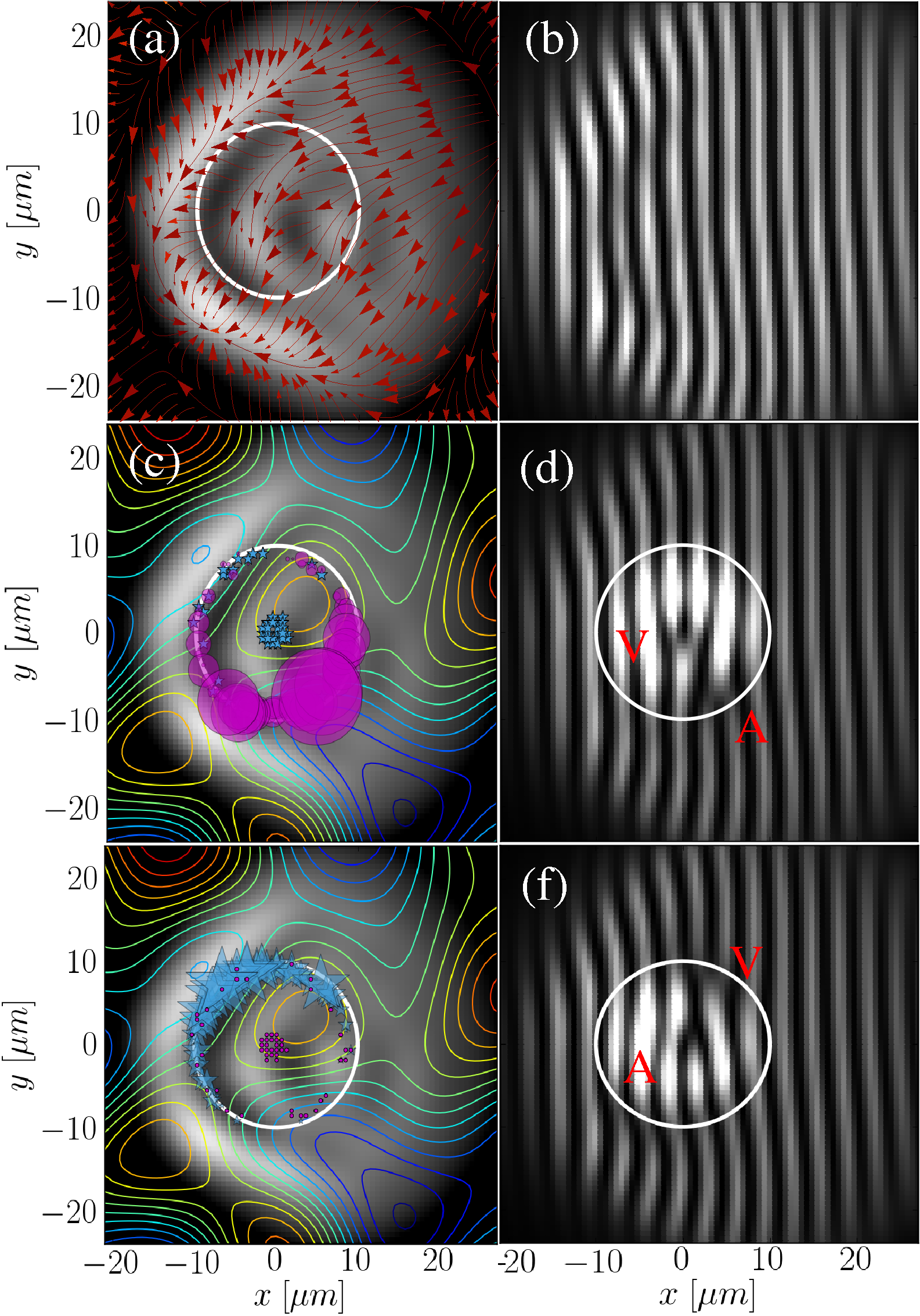}
\end{center}
\caption{(Color online) Simulated profile and supercurrents of the
  steady state OPO signal before the arrival of the probe (a) and
  associated interference fringes (b). Location of antivortices (dots
  (c)) and vortices (stars (e)) at the arrival ($t=0$~ps) of a vortex
  (stars (c)) or an antivortex (dots (e)) probe, for 1000 realisations
  of the random relative phase between pump and probe,
  $\Phi_{rdm}$. The size of dots in (c) (stars in (e)) is proportional
  to the number of times the antivortices (vortices) appear in that
  location. Panel (d) ((f)) shows single shot interference fringes
  relative to the plot in (c) ((e)). Contour-level lines in (c) and
  (e) represent the photonic disorder $V(\vect{r})$.  The white circle
  represents the edge of the probe.}
\label{fig:onere}
\end{figure}

However, if the extension of the probe carrying a vortex with charge
$m=+1$ is smaller than the size of the vortex-free OPO signal,
continuity of the polariton wavefunction requires that necessarily an
antivortex with charge $m=-1$ has to form at the edge of the probe
(see Fig.~\ref{fig:onere}).
In this Letter, we demonstrate that `unintended' antivortices do
appear in the signal at the edge of the imprinting vortex probe and
explain, both theoretically and via experiments, the origin of the
deterministic behaviour of the antivortex onset and dynamics. In
particular, we show where antivortices are more likely to appear in
terms of the supercurrents of the imprinting probe and the ones of the
underlying OPO. In addition, our study reveals that the onset of
vortices in polariton superfluids does not require a LG imprinting
beam, but instead vortex-antivortex (V-AV) pairs can be also generated
when counter-propagating currents are imposed, similarly to what
happens in normal (classical) fluids.
%
%Here, however, single-valueness of the superfluid wavefunction
%requires the vortices to be quantised.

Crucially, via numerical simulations, we elucidate the reason why an
experimental average over many shots allows detecting a vortex by
direct visualisation in density and phase profiles. Recently, it has
been suggested by stochastic simulations~\cite{wouters10} that
vortices in non-resonantly pumped polariton condensates undergo a
random motion which will hinder their direct detection, unless they
are close to being pinned by the stationary disorder potential and
thus follow a deterministic trajectory~\cite{lagoudakis10}.
In the case considered here of a superfluid generated by the the OPO,
we can instead explain a deterministic dynamics of the V-AV pair in
terms of the OPO steady state supercurrents, which determine a unique
trajectory for the pair, allowing their observation in multi-shot
measurements.
%
%%%%%%%%%%%%%%%%%%%%%%%%%%%%%%%%%%%%%%%
%Model
%%%%%%%%%%%%%%%%%%%%%%%%%%%%%%%%%%%%%%%
\paragraph*{Model}
The generalised Gross-Pitaevskii equations,
\begin{equation}
  i\partial_t \begin{pmatrix} \psi_X \\ \psi_C \end{pmatrix}
  = \begin{pmatrix} 0 \\ F \end{pmatrix} + \left[\hat{H}_0
    + \begin{pmatrix} g|\psi_X|^2& 0 \\ 0 &
      V\end{pmatrix}\right] \begin{pmatrix} \psi_X
    \\ \psi_C \end{pmatrix} \; ,
\label{eq:model}
\end{equation}
for coupled cavity and exciton fields $\psi_{C,X} (\vect{r},t)$ with
pump and decay~\cite{whittaker2005_b} model the OPO dynamics
($\hbar=1$).
The exciton-exciton interaction induces a non-linear dynamics of lower
(LP) and upper (UP) polaritons, the eigenstates of the
non-interacting Hamiltonian:
\begin{equation}
  \hat{H}_0 = \begin{pmatrix} \omega_{X} - i \kappa_X & \Omega_R/2
    \\ \Omega_R/2 & \omega_C-\nabla^2/(2m_C) - i
    \kappa_C \end{pmatrix} \; .
\end{equation}
The cavity (exciton) field decays with rate $\kappa_C$ ($\kappa_X$)
and is replenished by a continuous wave (cw) laser, $F_p(\vect{r},t) =
\mathcal{F}_{f_p,\sigma_p} (r) e^{i (\vect{k}_p \cdot \vect{r} -
  \omega_p t)}$, with a top-hat profile~\footnote{See
  Ref.~\cite{marchetti10} for details and choice of
  parameters.}. Above a pump strength threshold, the system is driven
into the OPO regime where signal and idler states (with energies
$\omega_{s,i}$ and wavevectors $\vect{k}_{s,i}$) get exponentially
populated. In addition, the OPO is probed by an extra pulsed laser
$F_{pb}(\vect{r},t)$. As single shot measurements would give a too low
signal to noise ratio, an average is performed over many pulsed
experiments taken always for the same OPO conditions. What differs at
each probe arrival is the random relative phase $\Phi_{rdm}$ between
pump and probe,
\begin{equation}
  F(\vect{r},t) = F_{p}(\vect{r},t) + F_{pb}(\vect{r},t)
  e^{i\Phi_{rdm}} \; ,
\end{equation}
with $\Phi_{rdm}$ uniformly distributed between $0$ and $2\pi$.

\begin{figure}
\begin{center}
\includegraphics[width=1.0\linewidth,angle=0]{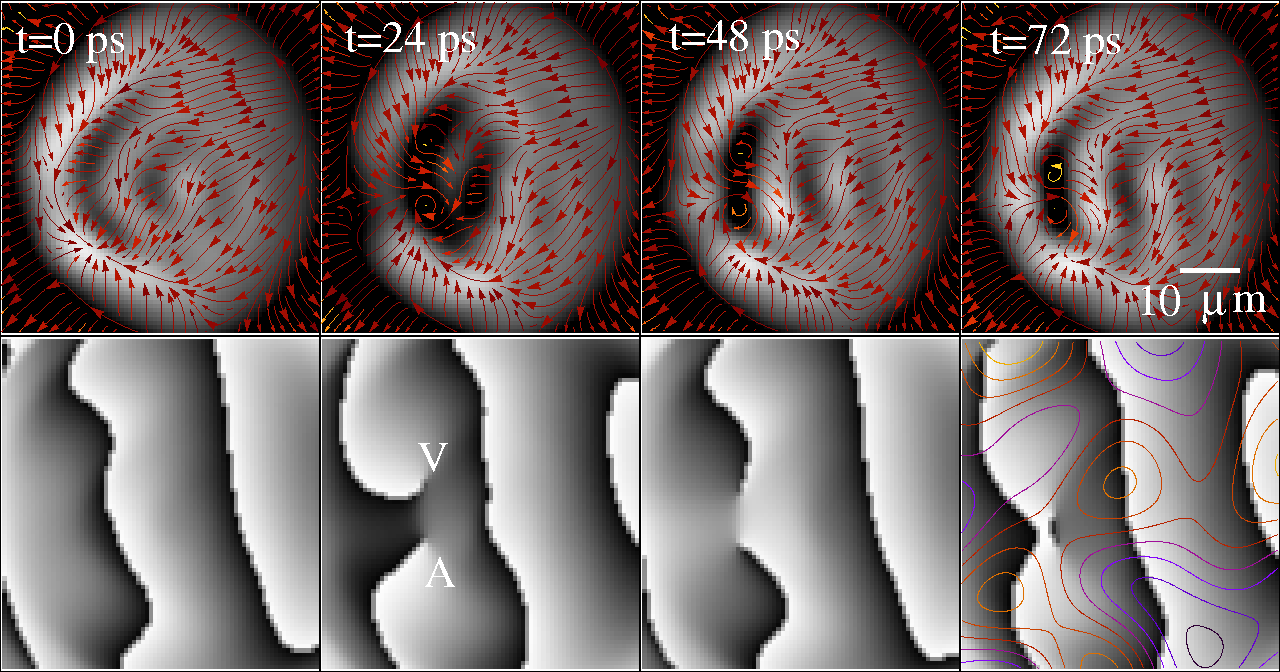}
\end{center}
\caption{(Color online) Simulated time evolution of signal after the
  arrival of the vortex probe, averaged over 1000 realisations of the
  random phase $\Phi_{rdm}$, $\langle \psi_{C}^s (\vect{r},t)
  \rangle_{\Phi_{rdm}}$ --- spatial profile (top) and phase
  (bottom). Contour-level lines in the last panel represent the
  photonic disorder.}
\label{fig:aver}
\end{figure}
As already shown in
Refs.~\cite{sanvitto10,krizhanovskii10,marchetti10}, vortices with
charge $m=\pm1$ can be imprinted in the OPO signal and idler, by
adding a LG pulsed probe:
\begin{multline}
  F_{pb}(\vect{r},t) = f_{pb} |\vect{r}-\vect{r}_{pb}|
  e^{-|\vect{r}-\vect{r}_{pb}|^2/(2\sigma^2_{pb})} e^{i m \varphi
    (\vect{r})} \\
  \times e^{i (\vect{k}_{pb} \cdot \vect{r} - \omega_{pb} t)}
  e^{-(t-t_{pb})^2/(2\sigma^2_{t})} \; ,
\label{eq:probe}
\end{multline}
where the probe momentum $\vect{k}_{pb}$ and energy $\omega_{pb}$ are
resonant with, e.g., the OPO signal state. Here, the pulse lasts
$2$~ps only. The azimuthal angle $\varphi (\vect{r})$ winds from $0$
to $2\pi$ around the vortex core $\vect{r}_{pb}$.
Finally, to mimic the experimental conditions, we include
in~\eqref{eq:model} a static photonic disorder potential
$V(\vect{r})$, with $\langle V(\vect{r}) \rangle =0$ and $\langle
V(\vect{r}) V(\vect{r}') \rangle = \sigma_d^2
e^{-|\vect{r}-\vect{r}'|^2/2 \ell_d^2}$ ($\ell_d \simeq 20\ \mu$m and
$\sigma_d \simeq 0.1$\ meV).

\begin{figure*}
\begin{center}
\includegraphics[width=0.75\linewidth,angle=0]{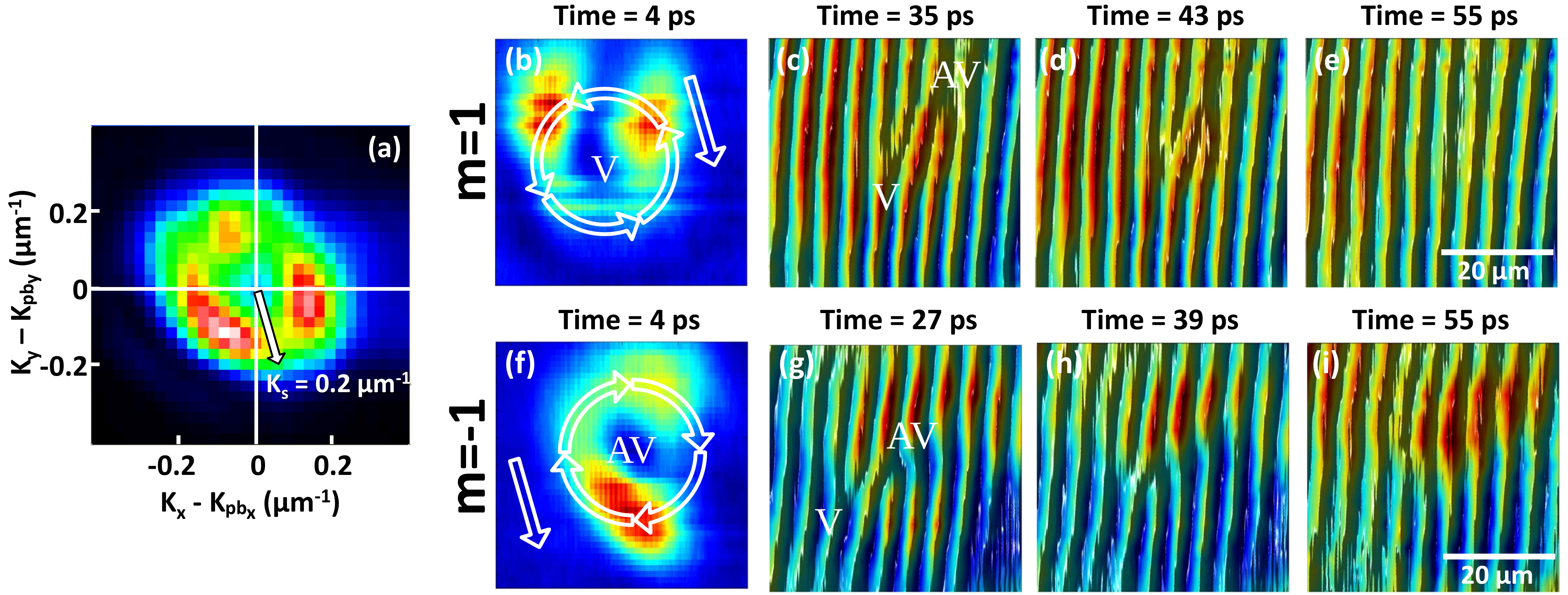}
\end{center}
\caption{(Color online) Measured momentum distribution of the vortex
  probe vs. $\vect{k}-\vect{k}_{pb}$ (a). The arrow indicates the
  signal momentum $\vect{k}_s-\vect{k}_{pb}$. Panel (b) ((f)) shows
  the real space emission of the signal $4$~ps after the $m=1$
  ($m=-1$) LG probe arrival. Panels (c--e) (panels (g--i)) compile,
  through the interference patterns, the time evolution of an
  imprinted $m=1$ vortex ($m=-1$ antivortex) and its associated $m=-1$
  antivortex ($m=1$ vortex). The straight arrow in (b,f) represents
  the signal current direction in the probe reference frame, while the
  probe current winds anti-clockwise for $m=1$ (clockwise for
  $m=-1$). The `unintended' antivortex (vortex) appears in (c) ((g))
  at the edge of the probe where the signal and probe currents are
  anti-parallel. V-AV pair dynamics can be followed in both cases for
  about $30$~ps, after which they annihilate.}
\label{fig:expe1}
\end{figure*}
We solve numerically Eq.~\eqref{eq:model} on a 2D grid by using a
5$^{\text{th}}$-order adaptive-step Runge-Kutta algorithm. We first
find the steady state stationary conditions for OPO emission
($f_{pb}=0$) and plot the OPO signal profile $|\psi_{C}^{s}
(\vect{r},t)| e^{i\phi_{C}^{s} (\vect{r},t)}$ by, e.g., filtering in a
cone around the signal momentum. In addition to the spatial profile,
$|\psi_{C}^{s} (\vect{r},t)|$, we also evaluate the supercurrents
$\nabla \phi_{C}^{s} (\vect{r},t)$ --- see
Fig.~\ref{fig:onere}(a). Note that the presence of the photonic
disorder does not change qualitatively our results. Its role is to
break the $y \mapsto -y$ symmetry left by the pump with
$\vect{k}_p=(k_{px},0)$ and to change accordingly the supercurrents.
Further, we simulate the dynamics following the arrival of a vortex
probe~\eqref{eq:probe} at $t=0$~ps for 1000 realisations of
$\Phi_{rdm}$ and then average
%the complex wavefunctions over such
%realisations at fixed time and space, 
$\langle |\psi_{C}^{s} (\vect{r},t)| e^{i\phi_{C}^{s}
  (\vect{r},t)}\rangle_{\Phi_{rdm}}$.

%%%%%%%%%%%%%%%%%%%%%%%%%%%%%%%%%%%%%%%
%Experimental setup
%%%%%%%%%%%%%%%%%%%%%%%%%%%%%%%%%%%%%%%
\paragraph*{Experimental setup}
The sample studied is a $\lambda/2$ AlAs microcavity with a single
GaAs quantum well placed at the anti-node of the mirror-confined
cavity field, giving a Rabi splitting of $4.4$~meV -- for details on
the sample see Ref.~\cite{perrin05}. Maintaining the sample at $10$~K,
a cw Ti:Sapphire laser, $F_p(\vect{r},t)$, resonantly pumps polaritons
at $1.5283$~eV and $k_{px}=1.4~\mu$m$^{-1}$. Above a threshold, the
system enters the steady state OPO regime. For a typical pump power of
$450$~mW, the signal emits at $1.5268$~eV, $1$~meV blue-shifted from
the LP bare dispersion. We filter the emission in $k$-space around the
signal momentum.
%to eliminate the pump contribution. 
A streak camera follows the evolution of the signal after the arrival
of a $2$~ps-long probe pulse from a second Ti:Sapphire laser,
$F_{pb}(\vect{r},t)$, resonant with the OPO signal (typical power
$3~\mu$W), and an average over millions of shots is performed.
The probe LG profile is generated by shining a Gaussian beam through a
hologram with a fork-like dislocation on its fringe
pattern. Interference images between the signal and an expanded,
constant phase, region of the signal are obtained in a Mach-Zehnder
interferometer.

%%%%%%%%%%%%%%%%%%%%%%%%%%%%%%%%%%%%%%%
%Results
%%%%%%%%%%%%%%%%%%%%%%%%%%%%%%%%%%%%%%%
\paragraph*{Results}
We discuss here the results obtained in the numerical simulations and
later in the experiments. In both cases, we choose OPO conditions such
to give a vortex-free signal (see Fig.~\ref{fig:onere}(a,b)).
Nevertheless, the simultaneous presence of pump, signal and idler
emitting at different momenta, as well as the photonic disorder,
implies that the OPO steady state is characterised by currents
carrying polaritons from gain- to loss-dominated regions. In
Fig.~\ref{fig:onere}(a), the signal currents have a dominant component
pointing leftwards and an equilibrium position where all currents
point inwards at around $(-8,-14)$~$\mu$m.

In single shot simulations of Fig.~\ref{fig:onere}(d,f) (one
realisation of the phase $\Phi_{rdm}$), we find that if the probe is
positioned well inside the OPO signal (e.g.,
$\vect{r}_{pb}=(0,0)$~$\mu$m), then the imprinting of a vortex $m=+1$
(antivortex $m=-1$) at $t=0$~ps forces the system to generate, at the
same time, an antivortex $m=-1$ (vortex $m=+1$) at the edge of the
probe. This is a consequence of the continuity of the polariton
wavefunctions: If the signal OPO phase is homogeneous and vortex-free
before the arrival of the probe, then imposing a topological defect,
i.e., a branch cut, on the signal phase at the probe core, requires
the branch cut to terminate where the phase is not imposed by the
probe any longer and has to continuously connect to the freely chosen
OPO signal phase, i.e. at the edge of the probe.
Note that OPO parametric scattering processes constrain the sum of
signal and idler phases to the phase of the laser pump by
$2\phi_p=\phi_s+\phi_i$. Thus, at the same positions where the V-AV
pair appears in the signal, an AV-V pair appears in the idler, so that
locally the phase constraint described above is satisfied. This agrees
with the experiments in~\cite{krizhanovskii10}, though there only a
single V (AV) in the signal (idler) could be detected, because the
signal size was comparable to the probe one.

Different relative phases $\Phi_{rdm}$ cause the antivortex (vortex)
to appear in different locations around the vortex (antivortex)
probe. However, on 1000 realisations of the random phase uniformly
distributed between $0$ and $2\pi$, we observe that the antivortices
(vortices) are more likely to appear on positions where the current of
the steady state OPO signal before the probe arrival and the probe
current are opposite. For example, for the $m=+1$ ($m=-1$) probe of
Fig.~\ref{fig:onere}(c) (Fig.~\ref{fig:onere}(e)), the current
constantly winds anti-clockwise (clockwise), therefore, comparing with
the signal current of Fig.~\ref{fig:onere}(a), the two are
anti-parallel in the bottom right (top left) region on the probe edge,
region where is very likely that an antivortex (vortex) is
formed. Note also that the onset of antivortices (vortices) privileges
regions where the steady OPO signal has a minimal intensity. Finally,
wavefunction continuity arguments allows the formation of additional
V-AV pairs on the probe edge, but are however rare events.

\begin{figure}
\begin{center}
\includegraphics[width=0.65\linewidth,angle=0]{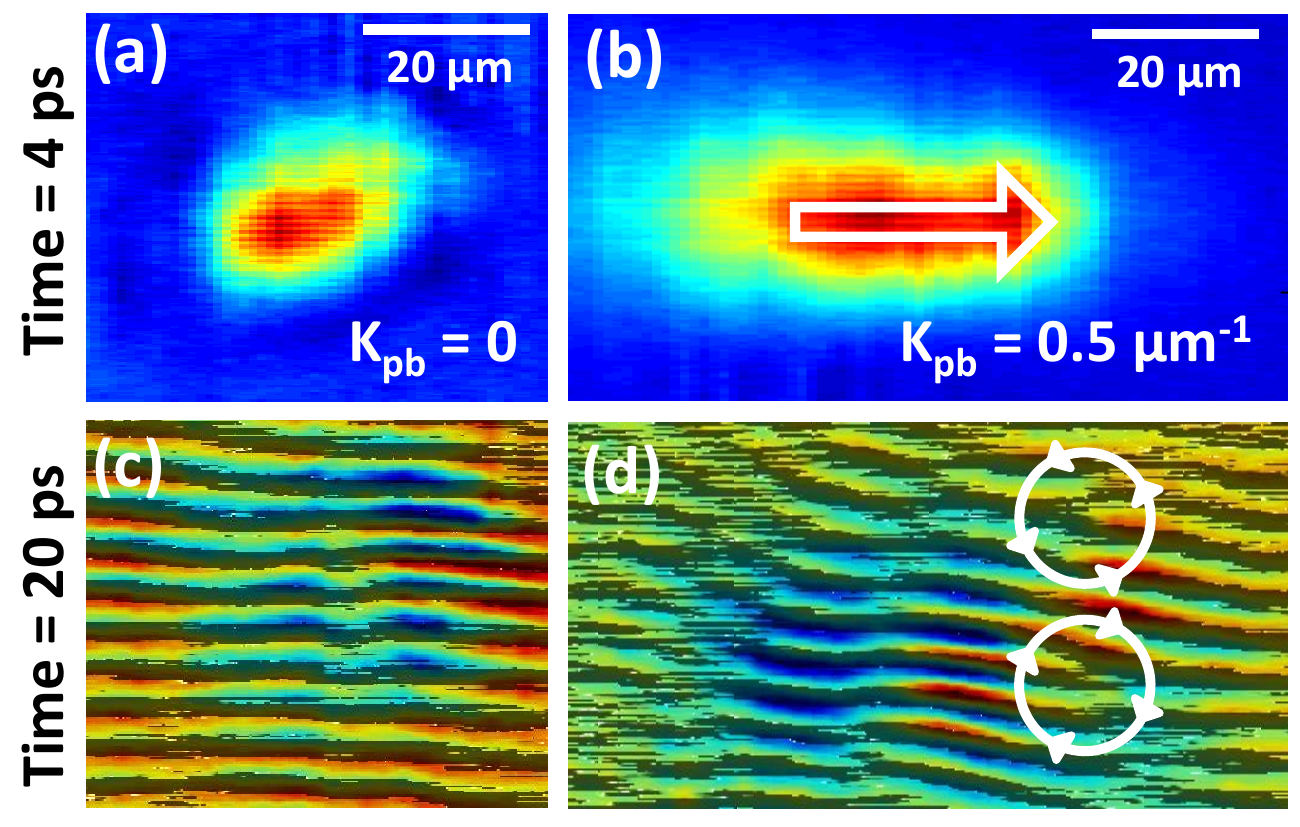}
\end{center}
\caption{(Color online) Gaussian probe at rest $\vect{k}_{pb}=0$ (a)
  and moving $\vect{k}_{pb}\ne 0$ (b) shined on the vortex-free OPO
  signal. The measured emission shows that no V-AV pair are created if
  the probe is at rest (c), while a pair appears after about $15$~ps
  for a moving probe (d).}
\label{fig:expe2}
\end{figure}
By averaging the 1000 images obtained at the probe arrival ($t=0$~ps),
e.g., in Fig.~\ref{fig:onere}(c), neither the imprinted vortex nor the
antivortex can be detected (see first panel of Fig.~\ref{fig:aver}):
Both phase singularities are washed away by averaging the differently
positioned branch-cuts. However, the steady state signal currents push
the V and AV, initially positioned in different locations, towards the
same equilibrium position where all currents point inwards. Thus,
while at $t=0$~ps, on average there is no V-AV pair, after $\sim
10$~ps, both V and AV appear and last $\sim 75$~ps (see
Fig.~\ref{fig:aver}), till they eventually annihilate.

The theoretical predictions are borne out by the experimental
observations. In order to confirm the role played by the relative
currents between the probe and the OPO signal on the appearance of the
`unintended' antivortex (vortex), we inject the vortex (antivortex)
probe with a finite momentum with respect to that of the signal. In
Fig.~\ref{fig:expe1}(a) we plot the momentum distribution of the probe
as a function of $\vect{k}-\vect{k}_{pb}$.
%, indicating with a straight arrow the value of
%$\vect{k}_s-\vect{k}_{pb}$. 
Thus, in the reference frame of the probe, the OPO signal has a
definite homogeneous current (straight arrow in
Figs.~\ref{fig:expe1}(b,f)), while the vortex (antivortex) probe has
anti-clockwise (clockwise) winding constant
currents. Figure~\ref{fig:expe1}(b) ((f)) shows the real space
emission of the signal $4$~ps after the $m=1$ ($m=-1$) LG probe
arrival. Images are taken by subtracting the steady state OPO.
Also, as in Ref.~\cite{sanvitto10}, in our experiments, the gain
triggered by the probe in the signal (and lasting around $25$~ps)
hinders the observation of the underlying signal dynamics, so that in
Fig.~\ref{fig:expe1} we show the signal evolution after the decay of
this extra population.
According to the previous analysis of Fig.~\ref{fig:onere}, we can
therefore predict the location of the `unintended' antivortex (vortex)
in Fig.~\ref{fig:expe1}(c) (Fig.~\ref{fig:expe1}(g)), namely where the
signal and probe currents are anti-parallel.
In particular, in Fig.~\ref{fig:expe1}(c) the antivortex appears on
the opposite side of the vortex in Fig.~\ref{fig:expe1}(g).
Despite the many-shot average, the dynamics of V-AV pairs can be
experimentally followed for about $30$~ps (Figs.~\ref{fig:expe1}(c--e)
and~(g--i)), thereafter the pair eventually annihilates.

Finally, we show that it is possible to create a V-AV pair with just a
Gaussian probe, when there is a difference in the signal and probe
currents. To this end, we shine a Gaussian pulsed beam either at rest
with respect to the OPO signal, $\vect{k}_{pb}=\vect{k}_s \simeq 0$,
or moving $\vect{k}_{pb}\ne \vect{k}_s$.  No pair appears in the first
case, while in the second, a V-AV pair appears on opposite sides of
the probe edge --- see Fig.~\ref{fig:expe2}.
Note that circulation is as expected, anti-clockwise (clockwise) for
the vortex (antivortex) on the upper (lower) side of the probe.

To conclude, the mechanism for V-AV pair formation reported here
differs from the V-AV binding-unbinding associated to the
Berezinskii-Kosterlitz-Thouless phase transition, recently adopted to
interpret the V-AV observation in non-resonantly pumped
polaritons~\cite{roumpos10}. In our case, the pair onset can be
explained in terms of OPO and probe relative currents, a simple
mechanism which does not require resorting to phase fluctuations
induced by the pump.

%%%%%%%%%%%%%%%%%%%%%%%%%%%%%%%%%%%%%%%%%%%%%%%%%
%%%%%%%%%%%%%%%%%%%%%%%%%%%%%%%%%%%%%%%%%%%%%%%%%
\acknowledgments G.T. acknowledges financial support from FPI
scholarship and F.M.M.  from the Ram\'on y Cajal program. Work
supported by the Spanish MEC (MAT2008-01555, QOIT-CSD2006-00019), CAM
(S-2009/ESP-1503) and FP7 ITN "Clermont4" (235114).
%%%%%%%%%%%%%%%%%%%%%%%%%%%%%%%%%%%%%%%%%%%%%%%%%

%\bibliography{biblio}

\newcommand\textdot{\.}

\end{document}